\documentclass[showpacs,preprintnumbers,prl,amsmath,amssymb,superscriptaddress, twocolumn]{revtex4}
\usepackage{graphicx}
\usepackage{endnotes}
\usepackage{bm}
\usepackage{epsfig}

\newcommand{\CaSn}{Ca$_3$Ir$_4$Sn$_{13}$}
\newcommand{\Tchimax}{$T^*$}
\begin{document}


\title{Enhancement of the Superconducting Transition Temperature with Hydrostatic Pressure in \CaSn\ Single Crystals}


\author{Swee~K.~Goh}
\email{skg27@cam.ac.uk}
\author{Lina~E.~Klintberg}
\affiliation{Cavendish Laboratory, University of Cambridge, J.J. Thomson Avenue, Cambridge CB3 0HE, United Kingdom}

\author{Patricia~L.~Alireza}
\affiliation{Cavendish Laboratory, University of Cambridge, J.J. Thomson Avenue, Cambridge CB3 0HE, United Kingdom}
\affiliation{Department of Physics and Astronomy, University College London, Gower St., London WC1E 6BT,
United Kingdom}

\author{David A. Tompsett}
\altaffiliation{Current address: Department of Chemistry, University of Bath, Bath BA2 7AY, United Kingdom}
\affiliation{Cavendish Laboratory, University of Cambridge, J.J. Thomson Avenue, Cambridge CB3 0HE, United Kingdom}

\author{Jinhu~Yang}
\author{Bin~Chen}
\affiliation{Department of Physics, Graduate School of Science, Hangzhou Normal University, Hangzhou 310036, China}
\affiliation{Department of Chemistry, Graduate School of Science, Kyoto University, Kyoto 606-8502, Japan}

\author{Kazuyoshi~Yoshimura}
\affiliation{Department of Chemistry, Graduate School of Science, Kyoto University, Kyoto 606-8502, Japan}

\author{F.~Malte~Grosche}
\affiliation{Cavendish Laboratory, University of Cambridge, J.J. Thomson Avenue, Cambridge CB3 0HE, United Kingdom}

\date{May 19, 2011}


\begin{abstract}
We report high pressure magnetic susceptibility and electrical resistivity measurements on Ca$_3$Ir$_4$Sn$_{13}$ single crystals up to 60~kbar. These measurements allow us to follow the evolution of the superconducting critical temperature $T_c$, the resistivity anomaly temperature \Tchimax, the superconducting coherence length and the Fermi velocity under pressure. The pressure-temperature phase diagram constructed for Ca$_3$Ir$_4$Sn$_{13}$ shows a dome-shaped pressure dependence of $T_c$. The initial rise in $T_c$, which is accompanied by a decrease in \Tchimax, is consistent with a reduction in the partial gapping of the Fermi surface under pressure.

\end{abstract}


\maketitle

\section{Introduction}
In many narrow-band $d$- or $f$-electron compounds, the low temperature ground state emerges out of a tangle of alternatives, including various forms of structural, magnetic or charge order, superconductivity, and other, more exotic possibilities. Often, the energy scales associated with these low temperature states change very differently with lattice density \cite{Lonzarich05}, which motivates the use of high pressure experiments to bring out the state of interest for detailed examination. This approach has been particularly successful in the case of superconductivity, because the nature of the state which is suppressed to make way for a superconducting ground state may itself give valuable clues to the nature and origin of the superconducting pairing mechanism.

\CaSn\ features a superconducting ground state with $T_c=7$~K at ambient pressure \cite{Yang10, Espinosa80, Espinosa82}. At \Tchimax$\simeq45$~K, distinct anomalies are observed in the electrical resistivity and in the magnetic susceptibility. The origin of these anomalies is so far unidentified. They have been attributed to strongly temperature dependent magnetic fluctuations \cite{Yang10}. The resistivity follows a non-Fermi liquid temperature dependence, and upon applying high magnetic field Fermi liquid behaviour is restored \cite{Yang10}. 

The related compound Sr$_3$Ir$_4$Sn$_{13}$ is also a superconductor, with $T_c=5$~K \cite{Espinosa80}. Since Ca is smaller than Sr and they are isovalent, the substitution of Ca for Sr can be regarded as providing positive chemical pressure. Here, we apply hydrostatic pressure to \CaSn\ to investigate whether $T_c$ can be enhanced further, and to study the interplay between $T_c$ and \Tchimax.

Although \CaSn\ crystals were synthesized almost 30 years ago \cite{Espinosa80, Espinosa82}, few studies have been reported until very recently. In order to track both $T_c$ and \Tchimax\ with pressure, we measure both the magnetic susceptibility and the electrical resistivity. To our knowledge, this is the first comprehensive high pressure study of this compound.

\section{Experimental}
The single crystals of \CaSn\ used for this study were grown by a flux method \cite{Espinosa80, Espinosa82}. Four-wire electrical resistivity measurements were performed using a miniature piston-cylinder cell, which can fit into the Quantum Design physical property measurement system (PPMS-9). In addition, two Moissanite anvil cells were prepared for AC susceptibility measurements with a conventional mutual inductance method, in which a 10-turn microcoil \cite{Goh08, Alireza03, Klintberg10} is placed inside the gasket hole as the pickup coil. A 140-turn coil was placed above the gasket to provide the modulation field. Glycerin was used as the pressure transmitting fluid for the piston-cylinder cell and one of the anvil cells, and 4:1 methanol-ethanol mixture was used for the other anvil cell. Ruby fluorescence spectroscopy was employed to determine the pressure achieved in the anvil cell. The superconducting transition temperature of \CaSn, as determined using the anvil cell, varies linearly with pressure at low pressures with $(\text{d}T_c/\text{d}P)_{initial}\sim0.83$~K/GPa.  For resistivity measurements carried out in the piston-cylinder cell, we use $T_c$ of \CaSn\ as an \textit{in situ} pressure gauge \cite{note1}. 

\section{Results and Discussion}

\begin{figure}[!t]\centering
      \resizebox{8.9cm}{!}{
              \includegraphics{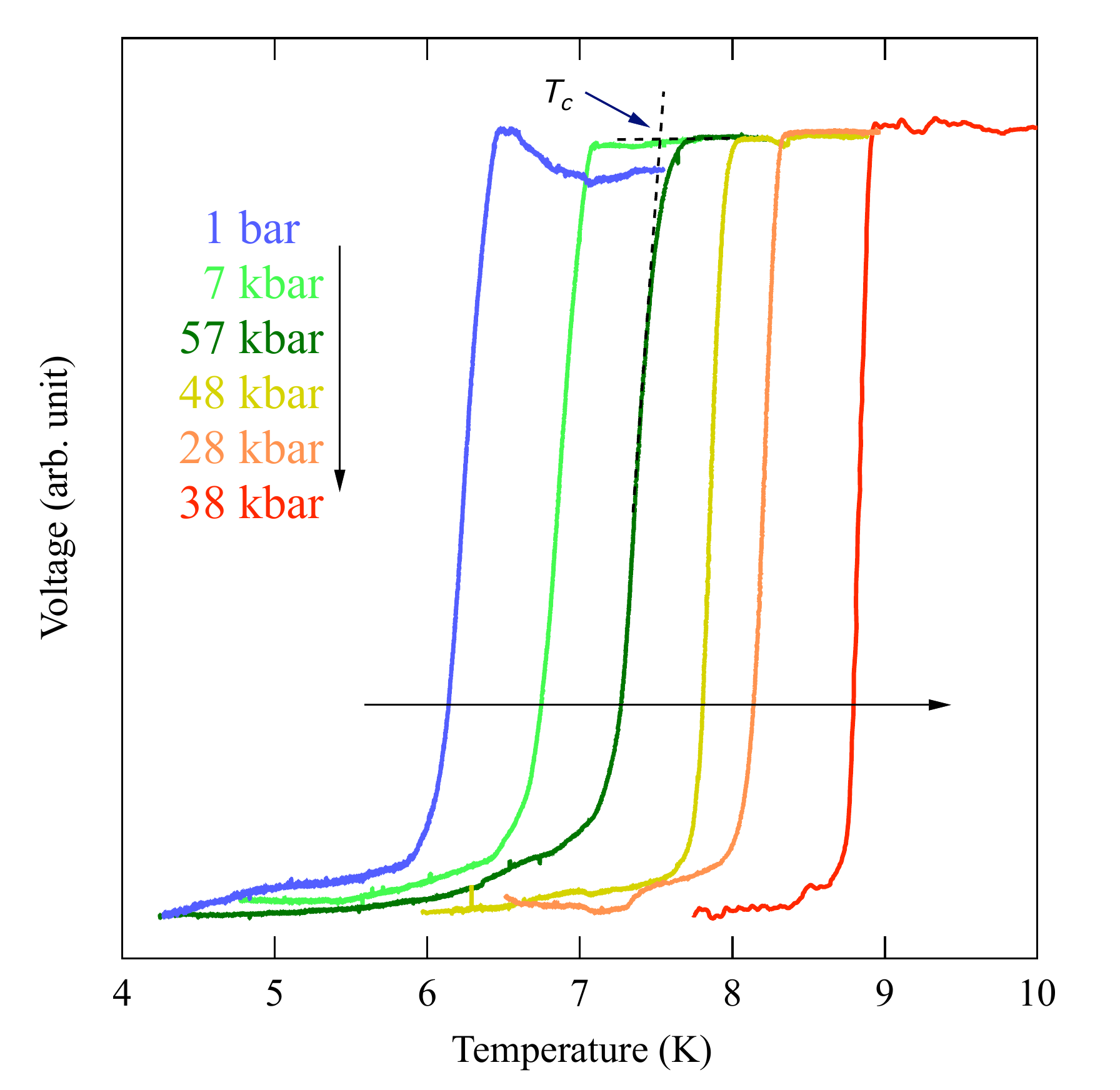}}               				
              \caption{\label{fig1} (Color online) Pressure dependence of the superconducting transition in \CaSn\ detected in Moissanite anvil cells using an AC mutual inductance technique.}
\end{figure}

Figure \ref{fig1} presents the temperature dependence of the AC susceptibility up to 57~kbar. The transition to the superconducting state is manifested by a clear drop in the voltage induced in the pickup coil. The transitions are all sharp, which demonstrates good hydrostaticity and high sample quality. As pressure is increased, $T_c$ first increases before reaching a maximum value of $\sim$8.9~K at about 40~kbar, giving a dome-shaped dependence of $T_c$ with pressure. Note that the ambient pressure $T_c$ extracted from magnetic measurements is 6.43~K, which is lower than the $\sim$7~K reported previously \cite{Yang10}. However, the pressure dependence of $T_c$ is clear and unambiguous.
\begin{figure}[!t]
  \centering
  \begin{minipage}[h]{40mm}\hspace{0mm}
    \resizebox{!}{80mm}{\includegraphics{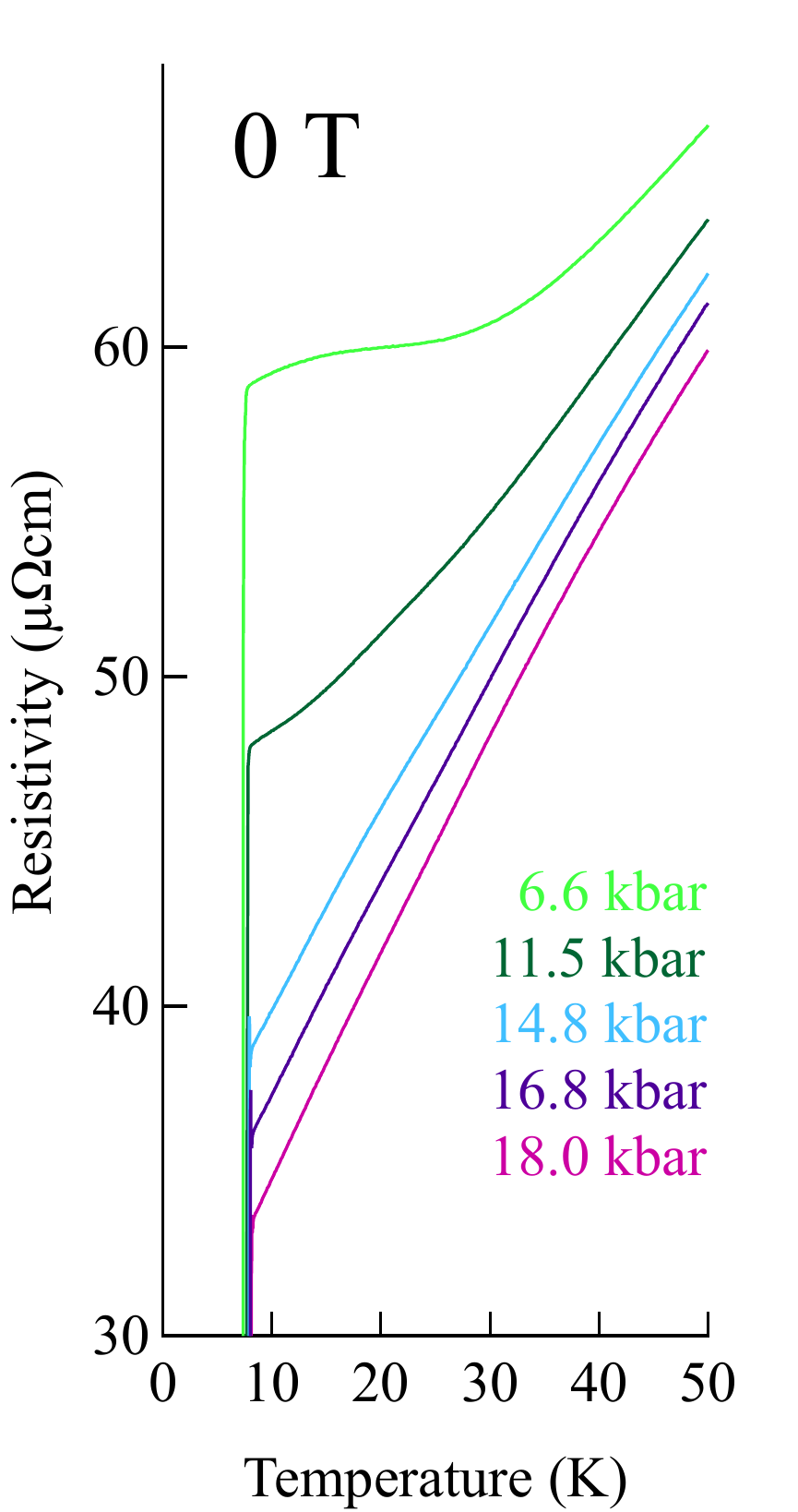}}
  \end{minipage}
    \begin{minipage}[h]{40mm}\hspace{-5mm}
    \resizebox{!}{80mm}{\includegraphics{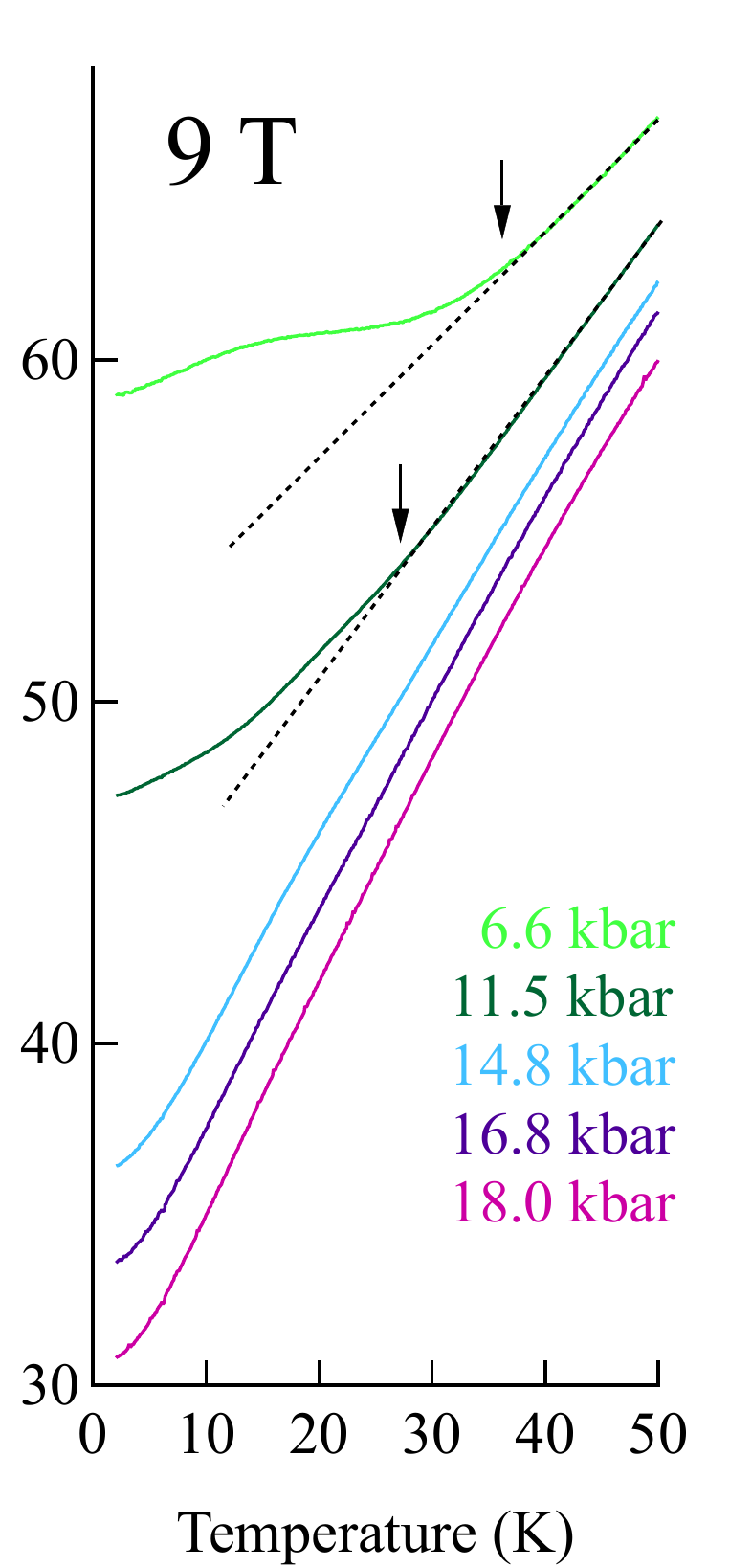}}
  \end{minipage}
 \caption{\label{fig2} (Color online) Temperature dependence of resistance measured at zero field (left) and 9~T (right). The arrows indicate \Tchimax (see text).}
\end{figure} 

As mentioned in the introduction, there is an interesting feature at \Tchimax, where a broad hump is observed in the resistivity curve.  At the onset temperature of this hump the magnetic susceptibility has been shown to peak \cite{Yang10}. The evolution of \Tchimax\ under pressure is important for understanding the superconductivity in this system, and since it is difficult to determine \Tchimax\ from high pressure measurements of the magnetic susceptibility, the resistivity data offer a precious opportunity to access this information. Figure \ref{fig2} shows the temperature dependence of the resistivity at zero field and 9~T up to 18~kbar. As shown in Figure \ref{fig2}, it is possible to assign \Tchimax\ to the resistivity traces at 6.6~kbar and 11.5~kbar. This feature is rapidly washed out under pressure, and it is no longer possible to determine \Tchimax\ conclusively for pressures higher than 11.5~kbar. Remarkably, while $T_c$ increases initially under pressure, \Tchimax\ decreases monotonically. 

In addition to shifting the $T_c$, the application of pressure lowers the residual resistivity $\rho_0$ significantly. This is apparent for the traces collected at 9~T where superconductivity is suppressed by magnetic field and the normal state is exposed down to 2~K, the base temperature of our measurement. As will be argued below, this drop in $\rho_0$ is positively linked to the initial increase in $T_c$.

\begin{figure}[!t]\centering
      \resizebox{8cm}{!}{
              \includegraphics{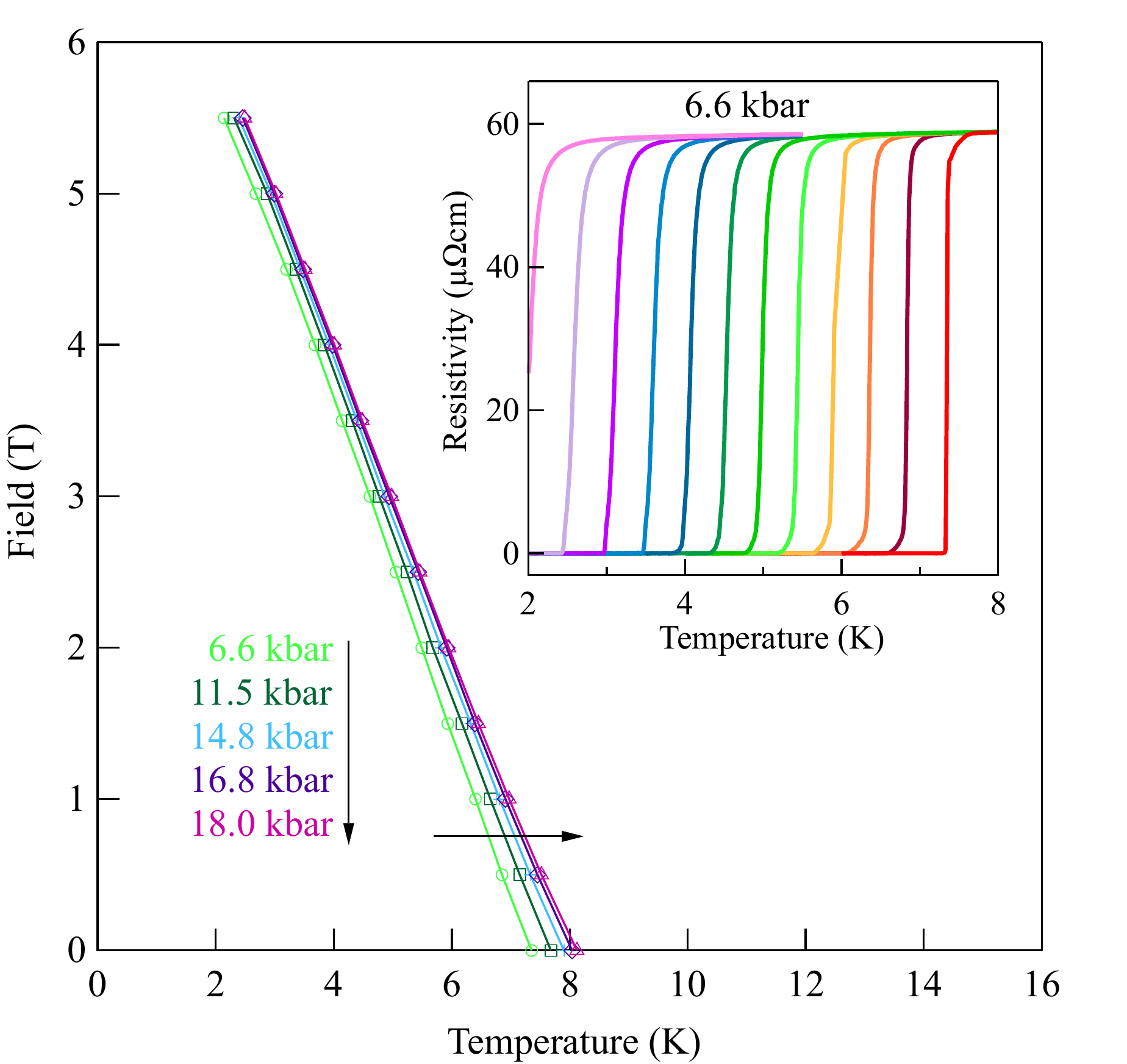}}               				
              \caption{\label{fig3} (Color online) Field dependence of $T_c$ at various pressures up to 18~kbar. Inset: Superconducting transition detected resistively at 6.6~kbar from 0~T to 5.5~T with 0.5~T step.}
\end{figure}

Figure \ref{fig3} (inset) depicts an example of resistivity measurements carried out at different magnetic fields; valuable information can be obtained by analyzing the magnetic field dependence of $T_c$. Within the temperature and field ranges accessible by our measurements, the variation of $T_c$ is approximately linear in field $H$ (Figure \ref{fig3}). As $T_c$ increases under pressure, the slope d$H$/d$T_c$ (or equivalently d$H_{c2}$/d$T$ where $H_{c2}$ is the upper critical field) decreases slightly.
The zero temperature upper critical field is proportional to $T_c\times(\text{d}H_{c2}/\text{d}T$) \cite{Tinkhambook}, and although both $T_c$ and d$H_{c2}$/d$T$ are pressure dependent, the results show that the product is pressure independent (up to 18~kbar). From this we can deduce that since the coherence length $\xi$ is inversely proportional to $H_{c2}$, then $\xi(T\rightarrow0)$ is also pressure independent.
BCS theory gives $\xi=a\hbar v_F/k_BT_c$ 
where $a$ is a constant whose value depends on the dimensionality of the system and the nodal structure of the superconducting gap \cite{Tinkhambook, Won94}. Provided that $a$ does not vary under pressure, our analysis suggests that $T_c$ is proportional to $v_F$.

The Fermi velocity $v_F$ can be written as $\hbar k_F/m^*$, which gives, for isotropic Fermi surfaces, $v_F=\hbar(3\pi^2n)^{1/3}/m^*$, where $k_F$, $m^*$ and $n$ are the Fermi wavevector, the effective mass and the carrier density, respectively \cite{Ashcroftbook}. Additionally, the Drude model gives $\rho^{-1}=ne^2\tau/m^*$ with $\tau$ as the scattering time \cite{Ashcroftbook}. The mean free path $l=\tau v_F$ at low temperature is mainly limited by the presence of impurities and defects, and is not altered by pressure. This is particularly true for our case, since the same single crystal is used throughout the experiment. Using the above equations
we can write $\rho^{-1}=(e^2l/3\pi^2\hbar)k_F^2$, where the mean free path can be treated as a constant.

\begin{figure}[!t]      \centering
 \resizebox{7.8cm}{!}{
              \includegraphics{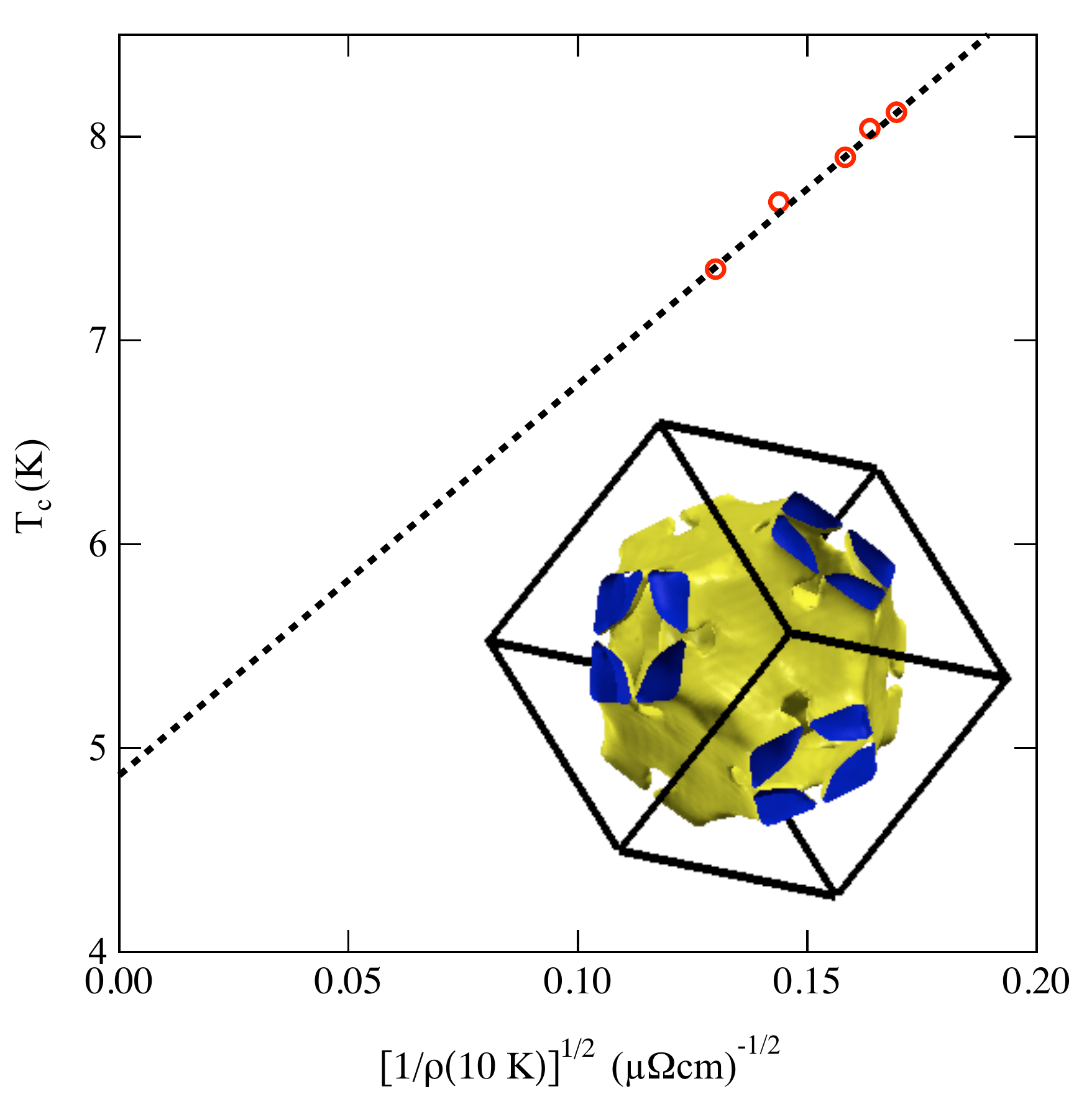}}               	       				
              \caption{\label{fig4} (Color online) The plot of $T_c$ against $\sqrt{1/\rho(10~\text{K})}$. The dotted line is a guide to the eye. Inset: Band 329 of the Fermi surface. The zone center is at the center of the cube.}
\end{figure}

Combining the BCS coherence length with $\rho^{-1}\propto k_F^2$, and $v_F\propto k_F/m^*$ we can write $T_c \propto (1/m^*)(1/\rho)^{1/2}$. Therefore, the drop in $\rho$ (Figure \ref{fig2}) can be attributed to an increase in $k_F$, which concomitantly results in an increase in $v_F$ and $T_c$. 
In Figure \ref{fig4}, we plot $T_c$ versus $\sqrt{1/\rho(10~\text{K})}$. Although a linear fit is able to describe the data well, a sizeable vertical offset of 4.9~K is observed. Note that our model above requires the data to extrapolate to the origin. Therefore, we conclude that $m^*$ is pressure dependent and with the values of $T_c$ and $\rho(10~\text{K})$, we can estimate the relative values of $m^*$. We find that $m^*$ increases monotonically under pressure, and $m^*(18~\text{kbar})/m^*(6.6~\text{kbar})\sim1.18$. The reason for this enhancement of $m^*$ is not clear, and therefore microscopic probes such as the de Haas-van Alphen effect will be useful to deepen our understanding of this compound. 

\begin{figure}[!t]\centering
     \resizebox{8.9cm}{!}{
              \includegraphics{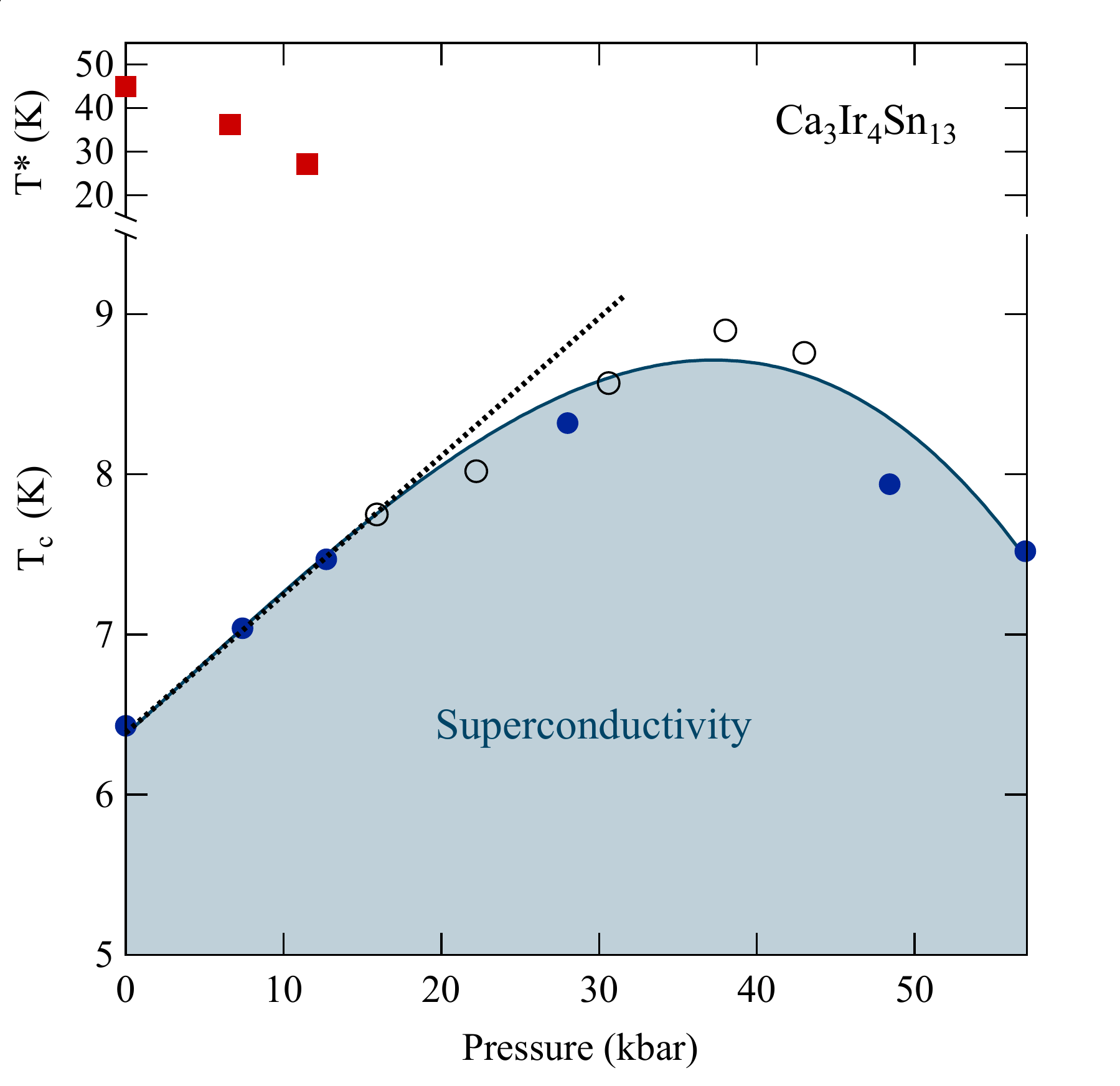}}               				
              \caption{\label{fig5} (Color online) Phase diagram showing the pressure dependence of $T_c$ ({\Large{$\circ${\normalsize{,}} $\bullet$}}) and \Tchimax ($\blacksquare$). For the mutual inductance measurement with the anvil cell, two pressure transmitting media were used -- methanol-ethanol mixture (in the ratio of 4:1) ({\Large{$\circ$}}) and glycerin ({\Large{$\bullet$}}). For the resistance measurement, only glycerin was used ($\blacksquare$). The dotted line is a guide to the eye, showing the linear variation of $T_c$ at low pressures.}
\end{figure}

In Figure \ref{fig5} we summarize the pressure dependence of $T_c$ and \Tchimax. $T_c$ follows a smooth dome-shaped variation with a maximum of $\sim$8.9~K at about 40~kbar, independent of the pressure transmitting media used for the experiment. \Tchimax, by contrast, decreases monotonically and extrapolates linearly to 0~K at $\sim$30~kbar. Since \Tchimax\ only exists for the part of the phase diagram where $\text{d}T_c/\text{d}p > 0$, it is tempting to conclude that these features are connected.

The suppression of \Tchimax\ is accompanied by a reduction in $\rho_0$. It is possible that for $T<\ $\Tchimax, a partial gapping of the Fermi surface occurs. The anomaly at \Tchimax, which is more pronounced at low pressures, is characteristic of the formation of a spin- or charge-density wave (SDW or CDW) [eg. \cite{Shelton86, Torikachvili08}]. In this scenario, partial gapping of the Fermi surface at \Tchimax\ abruptly reduces the density of states at the Fermi level, $g(E_F)$ and thereby the Pauli susceptibility, causing the observed drop in the magnetic susceptibility below \Tchimax\ \cite{Yang10}. Under pressure, the reduced gapping of the Fermi surface raises the effective carrier density and $g(E_F)$, which increases the electrical conductivity and boosts $T_c$. At pressures higher than 40~kbar, \Tchimax\ plays no role and no new carriers are added. The eventual decline in $T_c$ in this regime can be understood with the conventional scenario of lattice stiffening.

\CaSn\ crystallizes in \textit{Pm\=3n} space group which has a cubic symmetry. Although the observation of CDW and SDW states is well documented for low dimensional compounds, it is relatively rare for three dimensional systems. A well known example is the cubic spinel compound CuV$_2$S$_4$, which was found to display a CDW state at low temperatures \cite{Fleming81}. It is conceivable that the Fermi surface of \CaSn\ contains some low curvature sections which promote nesting. 
To gain further insight into the potential for CDW/SDW formation the electronic structure was calculated using the the Generalized Gradient Approximation \cite{Perdew96} with Wien2k \cite{Wien}. The experimental lattice coordinate of $a=9.7437$~\AA\ \cite{Yang10} was employed. $Rk_{max}=7$ and 40,000 $k$-points were used in a non-spin polarized calculation. The position of the 24$k$ Sn site (0, y, z), the only free internal coordinate, was minimised, resulting in (0, y = 0.3045, z = 0.1516). Six bands were found to cross the Fermi level. Each sheet of the Fermi surface was found to be three dimensional, but some showed the presence of significant flat sections that exhibit the potential for nesting. The flat sections of band 329 shown in the inset of Fig. \ref{fig4} are strong candidates for this. The potential importance of nesting in other three dimensional materials, e.g. MnSi and NbFe$_2$, has also been highlighted by electronic structure calculations \cite{Jeong04, Tompsett10}. The contribution of this band to the Lindhard susceptibility was also calculated and a strong peak was found at (1/2, 1/2, 1/2) at a value 21\% above that at $\Gamma$. This result provides a strong indication of the potential for nesting in this system, but the development of this instability will depend on other factors such as the q-dependent interaction parameter. Further results relating to these calculations will be reported elsewhere.

To summarise, in the cubic transition metal compound \CaSn\ an
unidentified electronic anomaly at $T^*$ is suppressed at a critical
pressure of about $30 ~\rm{kbar}$. This suppression is accompanied by
a distinct increase in the superconducting transition temperature,
which rises to a maximum of $8.9~\rm K$ at $40~\rm{kbar}$, producing a
broad superconducting dome in the temperature-pressure phase
diagram. In a series of high field experiments under pressure, we have
been able to extract the pressure dependence of the superconducting
coherence length, which can in a simplified model be linked to the
effective Fermi wavevector $k_F$. Our data indicate that $k_F$
initially increases with pressure. This observation is consistent with
the assumption that charge or spin density wave order sets in at
$T^*$, which gaps out parts of the Fermi surface and thereby reduces
$k_F$ and consequently the electronic density of states and
$T_c$. Removing this transition then boosts superconductivity.  A more
detailed examination of the $T^*$ anomaly is under way to clarify its
precise microscopic origin.

\textbf{Acknowledgement.} This work was supported by the EPSRC UK, Trinity College (Cambridge), Grant-in-Aid for Scientific Research from the JSPS (22350029), Global COE Program "International Center for Integrated Research and Advanced Education in Materials Science", Kyoto University, from the MEXT Japan. SKG acknowledges the Great Britain Sasakawa Foundation for travel grant and Kyoto university for hospitality.
  





\begin{thebibliography}{34}
\bibitem{note1} Since the ambient pressure $T_c$ extracted from AC susceptibility is lower, we used the published ambient pressure $T_c$ of 7~K \cite{Yang10},  and $(\text{d}T_c/\text{d}P)_{initial}$ to calculate the pressure achieved in the piston cylinder cell.
\end{thebibliography}
\end{document}